\documentclass[aps,prb,superscriptaddress,amsfonts,amsmath,amssymb,floatfix,reprint,longbibliography]{revtex4-2}

\usepackage{url}
\usepackage{bm}
\usepackage{graphicx}
\usepackage{amsmath}
\usepackage{amstext}
\usepackage{amssymb}
\usepackage{amsfonts}
\usepackage{amsbsy}
\usepackage{verbatim}
\usepackage{color}
\usepackage[colorlinks=true, urlcolor=blue, linkcolor=blue, citecolor=blue, pdftex]{hyperref}
\usepackage{multirow}
\usepackage{floatrow}
\usepackage{float}
\usepackage{tikz}
\usepackage{gensymb}
\usepackage{textcomp}
\usepackage{enumitem}
\usepackage[version=3]{mhchem}
\usepackage{afterpage}
\usepackage{pbox}
\usepackage{makecell}
\usepackage{dsfont}
\usepackage{physics}
\usepackage{siunitx}
\usepackage{fancyhdr}
\usepackage{stackengine,wasysym,scalerel}
\usepackage{latexsym}
\usepackage{cancel}
\usepackage{array, multirow, bigdelim, makecell, booktabs}
\usepackage[misc]{ifsym}
\usepackage[normalem]{ulem}
\usepackage{times}
\usepackage{comment}

\newcommand{\dagga}{{\phantom{\dagger}}}

\definecolor{darkblue}{HTML}{004D6B}
\definecolor{darkred}{HTML}{8c1515}
\definecolor{darkgreen}{HTML}{006400}
\usepackage{hyperref}
\hypersetup{
	colorlinks=true,
	urlcolor=darkred,
	citecolor=darkblue,
	linkcolor=darkred,
	breaklinks
}

\begin{document}

\title{Piercing the Dirac spin liquid: From a single monopole to chiral states}

\author{Sasank Budaraju}
\affiliation{Laboratoire de Physique Th\'eorique, Universit\'e de Toulouse, CNRS, UPS, France}
\affiliation{Department of Physics and Quantum Centre of Excellence for Diamond and Emergent Materials (QuCenDiEM), Indian Institute of Technology Madras, Chennai 600036, India}

\author{Yasir Iqbal}
\affiliation{Department of Physics and Quantum Centre of Excellence for Diamond and Emergent Materials (QuCenDiEM), Indian Institute of Technology Madras, Chennai 600036, India}

\author{Federico Becca}
\affiliation{Dipartimento di Fisica, Universit\`a di Trieste, Strada Costiera 11, I-34151 Trieste, Italy}

\author{Didier Poilblanc}
\affiliation{Laboratoire de Physique Th\'eorique, Universit\'e de Toulouse, CNRS, UPS, France}

\date{\today}

\begin{abstract}
The parton approach for quantum spin liquids gives a transparent description of low-energy elementary excitations, e.g., spinons and emergent
gauge-field fluctuations. The latter ones are directly coupled to the hopping/pairing of spinons. By using the fermionic representation of the 
$U(1)$ Dirac state on the kagome lattice and variational Monte Carlo techniques to include the Gutzwiller projection, we analyse the effect of 
modifying the gauge fields in the spinon kinematics. In particular, we construct low-energy monopole excitations, which are shown to be gapless 
in the thermodynamic limit. States with a finite number of monopoles or with a finite density of them are also considered, with different patterns 
of the gauge fluxes. We show that these chiral states are not stabilized in the Heisenberg model with nearest-neighbor super-exchange couplings, 
and the Dirac state corresponds to the lowest-energy {\it Ansatz} within this family of variational wave functions. Our results support the idea 
that spinons with a gapless conical spectrum coexist with gapless monopole excitations, even for the spin-1/2 case.
\end{abstract}

\maketitle

{\it Introduction.}
Quantum spin models on frustrated low-dimensional lattices represent a playground to investigate a variety of different phases of matter and 
the transitions among them~\cite{lacroix2011}. Even though a full characterization of their phase diagrams would require a finite-temperature 
analysis, in most cases the knowledge of the ground state and a few low-energy excitations is enough to obtain important information on the 
relevant (low-temperature) behavior. Still, achieving an accurate description of the exact ground state of frustrated spin models poses itself 
as a difficult task. Indeed, a faithful characterization can be obtained whenever (a sizable) magnetic order is present, since here the ground 
state is well approximated by a product state, with spins having well-defined expectation values on each site. By contrast, whenever magnetic 
order is significantly suppressed, or even absent, the ground-state wave function is much more elusive. The most complicated case is given by 
the so-called quantum spin liquids, where the elementary degrees of freedom are no longer the original spin variables, but emergent particles 
(spinons) and gauge fields (visons or magnetic monopoles)~\cite{savary2017}. The standard approach to describe spin liquids is through the 
parton construction, where spin operators are represented by using fermionic or bosonic particles; here, the original Hilbert space is enlarged 
and additional gauge fields are introduced~\cite{baskaran1988,arovas1988,affleck1988}. Thus, the resulting model describes fermions or bosons 
that interact through gauge fields on a lattice. A spin liquid corresponds to the deconfined phase of the resulting model, in which particles
(spinons) are free at low energies. In this case, the elementary excitations of the spin model are fractionalized, i.e., they are not integer 
multiples of those of the original constituents. By contrast, whenever the gauge fields lead to confinement, the spin liquid is unstable towards 
some symmetry-breaking phenomenon, most notably the establishment of valence-bond or magnetic order~\cite{read1990}. The analysis of these 
lattice gauge theories is not easy and requires non-perturbative methods~\cite{xu2019,song2019,song2020}, which also include a detailed 
examination of the symmetries of low-energy excitations. Still, some insight can be obtained from mean-field approaches~\cite{wen2002}, where 
gauge fields are frozen and fermions/bosons are free. From there, it is also possible to extract some information on the nature of the most 
relevant gauge fluctuations: whenever they are gapped (corresponding to a ${\mathbb Z}_2$ symmetry) the low-energy spectrum of the spinons is 
not qualitatively modified, leading to stable ${\mathbb Z}_2$ spin liquids~\cite{wen2002} (the most remarkable example being the Kitaev model 
on the honeycomb lattice~\cite{kitaev2006}). The situation is more delicate when the low-energy gauge fields are gapless (with $U(1)$ symmetry), 
since in this case they can spoil the mean-field properties of the spinon spectrum. In particular, monopoles proliferate and may give rise to 
a confined phase~\cite{polyakov1977}. Still, the presence of a sufficiently large number of massless fermions may screen the monopoles and 
prevent confinement~\cite{borokhov2002,hermele2004,xu2019}.

\begin{figure*}
\includegraphics[width=\columnwidth]{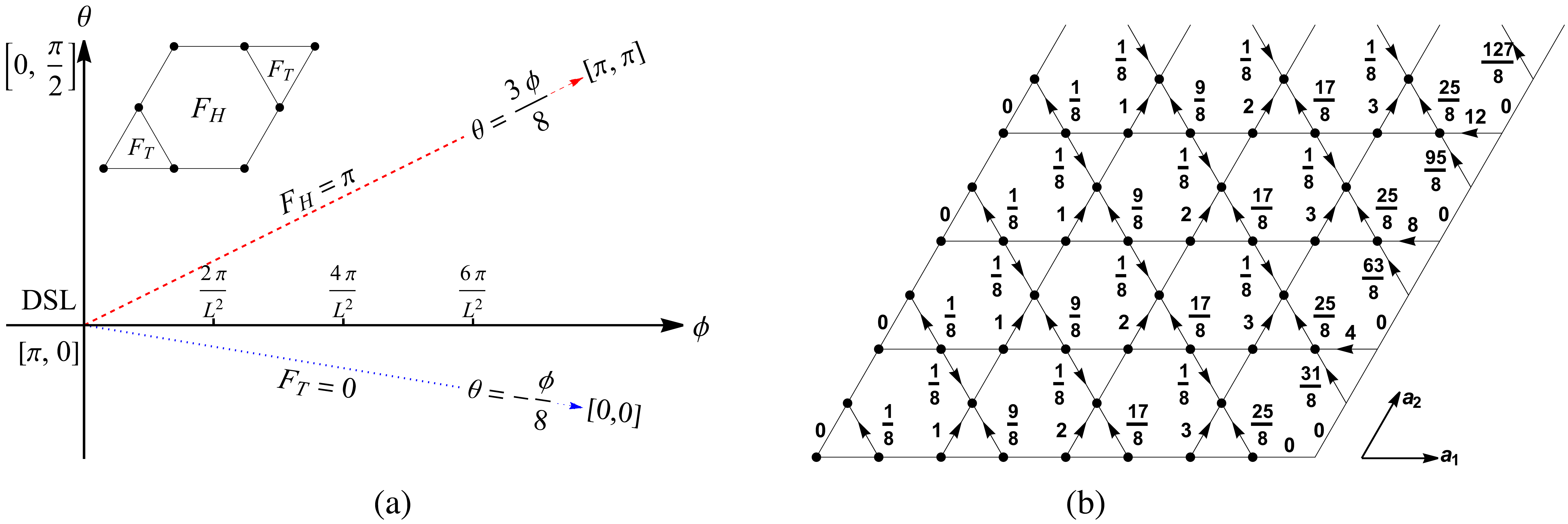}
\caption{\label{fig:fluxes} 
(a) The plane $(\phi,\theta)$ that defines the flux distribution in the unit cell considered in this work, shown in the inset. The hexagonal 
plaquette has a $F_{H}=\pi-2\theta+3\phi/4$ and two triangular ones have flux $F_{T}=\phi/8+\theta$. The $[\pi,0]$ Dirac state lies at the origin, 
the uniform state $[0,0]$ is obtained with $\theta=-\phi/8$ for $\phi=\pm \pi$, and the $[\pi,\pi]$ state with $\theta=3\phi/8$ and $\phi=2\pi$. 
The quantized values of $\phi$, obtained for a few monopoles are marked on the $x$-axis. (b) The complex argument $\alpha_{i,j}$, in units of 
$2\pi/L^2 = 2\pi/16$, of the hopping parameters $e^{\mathrm{i} \alpha_{i,j}}$ (for $i \to j$) and $e^{-\mathrm{i} \alpha_{i,j}}$ (for $j \to i$) 
of the fermionic Hamiltonian~\eqref{eq:aux_ham} that defines a single-monopole configuration (with $\theta=0$) on the $L=4$ cluster. Note that the hoppings on the last column break translational symmetry along the diagonally upwards direction.}
\end{figure*}


Among various possibilities, the nearest-neighbor $S=1/2$ Heisenberg antiferromagnetic model on the kagome lattice represents one of the most 
intriguing and important examples in which magnetic frustration may give rise to a non-magnetic ground state. The interest in this spin model 
was raised after the discovery of a number of compounds, where localized $S=1/2$ moments interact through a super-exchange mechanism in almost 
decoupled kagome layers. The most notable example is given by the so-called Herbertsmithite Cu$_3$Zn(OH)$_6$Cl$_2$~\cite{mendels2007,helton2007,devries2008}. 
Here, there is no evidence of magnetic order down to extremely small temperatures, thus suggesting the possibility that the ground state is 
indeed a quantum spin liquid~\cite{norman2016}. Triggered by these outcomes, a huge effort has been spent in the last years to clarify the actual 
nature of the ground state of the Heisenberg model on the kagome lattice.
From the theoretical side, exact diagonalizations of the Heisenberg model on small clusters highlighted the existence of a very unconventional low-energy spectrum, with an exceedingly large number of singlet states below the lowest triplet excitation \cite{lecheminant,lauchliED}. 
Early large-scale density-matrix renormalization group (DMRG) calculations suggested the existence of a gapped spin liquid~\cite{yan2011,depenbrock2012},
while variational Monte Carlo techniques, more recent DMRG and tensor network approaches, and pseudo-fermion functional renormalization group calculations supported a gapless spin liquid~\cite{ran2007,iqbal2013,he2017,liao2017,hering2019}. The variational approach has a very simple and elegant description within the fermionic parton representation; here, the free fermions have only 
kinetic terms (no pairing), defining peculiar magnetic fluxes piercing the unit cell (i.e., $\pi$-flux through hexagonal plaquettes and $0$-flux 
through triangular ones), thus leading to two Dirac points in the spinon spectrum~\cite{hastings2000,ran2007}. As a consequence, this {\it Ansatz} 
is dubbed as $[\pi,0]$ Dirac spin liquid. Finally, an accurate variational wave function is obtained by including the Gutzwiller projection, which 
imposes a single-fermion occupation on each lattice site~\cite{ran2007,iqbal2013}. 

Still, alternative scenarios have been proposed, the most intriguing ones suggesting the possibility that the ground state
is a (non-chiral) topological spin liquid \cite{mei2017} or a chiral spin 
liquid~\cite{messio2012,sun2022}, which break time-reversal and point-group symmetries~\cite{wen1989}. Originally, chiral spin liquids have been 
constructed in analogy to the fractional quantum Hall effect~\cite{kalmeyer1987}. However, the main difference with respect to the latter case is 
that time-reversal is spontaneously broken, leading to even more exotic phenomena~\cite{bieri2016}. Recently, different calculations suggested that 
chiral spin liquids may exist in extended Heisenberg models on the kagome lattice, e.g., adding super-exchange couplings at second or third neighbors,
multi-spin interactions, or Dzyaloshinskii-Moriya terms~\cite{he2014,gong2014,zhu2015,kumar2015,messio2017,wietek2015,gong2015,he2015,kiese2023,ferrari2023}. 
In addition, chiral spin liquids have been also analysed within mean-field approaches, in terms of both bosonic~\cite{messio2013,lugan2022} 
and fermionic partons~\cite{bieri2015,bieri2016}.

In this paper, we study the stability of the Dirac spin liquid wave function, which has been proposed to capture the correct ground-state properties 
of the nearest-neighbor Heisenberg model on the kagome lattice~\cite{ran2007,iqbal2013}, against chiral perturbations. We analyse the energetics of 
Gutzwiller-projected fermionic states that are obtained by adding non-trivial magnetic fluxes to the ones that define the Dirac wave function. 
In particular, we can independently (i) consider an additional flux (parametrized by $\phi$ and spread uniformly on the lattice) and/or (ii) 
redistribute the flux inside the unit cell (parametrized by $\theta$); hence, we assume that every unit cell has the same distribution of fluxes in 
the hexagonal and triangular plaquettes, see Fig.~\ref{fig:fluxes}. The flux through the triangular plaquettes is given by $F_{T}=\phi/8+\theta$, 
while the flux through the hexagonal ones is $F_{H}=\pi-2\theta+3\phi/4$, such that the total flux piercing the unit cell is $F_{C}=\pi+\phi$, 
the Dirac state being recovered with $\phi=\theta=0$. All calculations are performed on tori with $3\times L \times L$ sites by using variational 
Monte Carlo techniques to assess the properties of the Gutzwiller-projected states~\cite{becca2017}. On finite clusters, $\phi$ is quantized,
while $\theta$ may assume any value. A ``commensurate'' flux $\phi=2\pi/q$ requires a large super-cell that includes $q$ unit cells (assuming $q$ 
divides $L$) and implies a total flux multiple of $2\pi L $ on the whole torus. In addition to these standard cases, we also consider monopole 
configurations. A single monopole brings a $2\pi$ flux on the torus, thus leading to $\phi=2\pi/L^2$ on each unit cell; states with $N_{\rm mp}$ 
monopoles are then constructed by considering a flux density $\phi=2\pi N_{\rm mp}/L^2$. On the one hand, this allows us to study the energetics of 
a single monopole on finite clusters and its scaling in the thermodynamic limit; on the other hand, with monopole configurations, the stability of 
the Dirac state may be assessed for very small additional fluxes (i.e., much smaller than the minimal one accessible within the commensurate fluxes). 
The main outcome of this study is that the Dirac state is stable against chiral perturbations. Still, monopole excitations are gapless in the 
thermodynamic limit. We would like to emphasize that, since we work on tori, the analysis of the monopole energy cannot be directly connected to 
the scaling dimensions, as usually done within conformal-field theories, which consider a spherical geometry~\cite{borokhov2002,hermele2004,dupuis2021,he2022}.

{\it Model and methods.}
We study the Heisenberg model on the kagome lattice with nearest-neighbor super-exchange interaction $J>0$
\begin{equation}\label{eq:hamiltonian}
{\cal H} = J \sum_{\langle i,j \rangle} {\bf S}_i \cdot {\bf S}_j,
\end{equation}
where ${\bf S}_i=(S^x_i,S^y_i,S^z_i)$ is the spin-1/2 operator on a site $i$; periodic-boundary conditions are assumed on a cluster with 
$3 \times L \times L$ sites.

The variational wave functions are defined by
\begin{equation}\label{eq:psi}
|\Psi \rangle = {\cal P}_G |\Phi_0 \rangle,
\end{equation}
where $|\Phi_0 \rangle$ is the ground state of the auxiliary (non-interacting) Hamiltonian:
\begin{equation}\label{eq:aux_ham}
{\cal H}_0=\sum_{\langle i,j \rangle, \sigma} \chi{^\dagga_{i,j}} c^\dagger_{i,\sigma} c^\dagga_{j,\sigma} + \text{h.c.},
\end{equation}
where $c^\dagger_{i,\sigma}$ ($c^\dagga_{i,\sigma}$) creates (destroys) a fermion on site $i$ with spin $\sigma= \uparrow, \downarrow$;
$\chi^\dagga_{i,j}=\chi^{0}_{i,j}e^{\mathrm{i} \alpha_{i,j}}$ defines the hopping amplitude for nearest-neighbor sites $(i,j)$. The ``bare'' term
$\chi^{0}_{i,j} = \pm 1$ defines the $[\pi,0]$ flux pattern of the Dirac spin liquid, while the presence of $\alpha_{i,j} \ne 0$ allows us to 
consider $\theta \ne 0$ and/or $\phi \ne 0$ (including single- or multi-monopole states), see Fig.~\ref{fig:fluxes}. In addition, periodic- or 
anti-periodic-boundary conditions can be taken in ${\cal H}_0$. In practice, the auxiliary Hamiltonian is diagonalized and $|\Phi_0 \rangle$ is 
constructed as the Slater determinant of the lowest $N$ single-particle orbitals (where $N=3L^2$), which is well defined whenever there is a closed 
shell configuration, i.e., a finite-size gap between the $N$-th and the $(N+1)$-th levels. For commensurate fluxes, we adopt the Landau gauge, which 
implies a $q \times 1$ super-cell. By contrast, the single-monopole configuration requires a super-cell as large as the entire cluster (which remains 
the case also for multi-monopole configurations). A similar monopole construction has been discussed in Ref.~\cite{poilblanc1990} for the square lattice.
We remark that, whenever a single monopole is considered on top of the Dirac state, there is an exact degeneracy at the Fermi level (which is robust 
to changing the boundary conditions \cite{SUN}), with two levels per spin, i.e. four levels occupied by two fermions giving rise to 6 monopoles (3 singlets and 1 
triplet)~\cite{song2019,wietek2023}. We verified that any occupation of these levels gives the same variational energy. In this case, the unprojected 
state $|\Phi_0 \rangle$ does not correspond to a closed shell configuration and we use the single-particle orbitals obtained by the real-space 
diagonalization, without imposing any lattice symmetry. Then, monopole configurations do not correspond to specific $k$ points of the Brillouin zone.

Finally, ${\cal P}_G$ is the Gutzwiller projection onto the configuration space with one particle per site:
\begin{equation}\label{eq:gutz}
{\cal P}_G=\prod_i (n_{i,\uparrow} - n_{i,\downarrow})^2,
\end{equation}
where $n_{i,\sigma}=c^\dagger_{i,\sigma} c^\dagga_{i,\sigma}$. As a result, $|\Psi \rangle$ of Eq.~\eqref{eq:psi} defines a faithful variational 
wave function for the spin Hamiltonian~\eqref{eq:hamiltonian}. Standard Monte Carlo sampling based upon Markov chains is used to evaluate the 
variational energy~\cite{becca2017}. For the Hamiltonian \eqref{eq:hamiltonian}, the Dirac state has an energy per site Dirac state has an energy per site $e \approx -0.429$, which is higher than the best DMRG and tensor network estimates for the ground state, $e \approx -0.438$ \cite{depenbrock2012, liao2017}.  Still, this simple variational state may well capture the correct properties of the actual ground-state wave function, as suggested by recent DMRG calculations \cite{he2017}.
 
\begin{figure}
\includegraphics[width=\columnwidth]{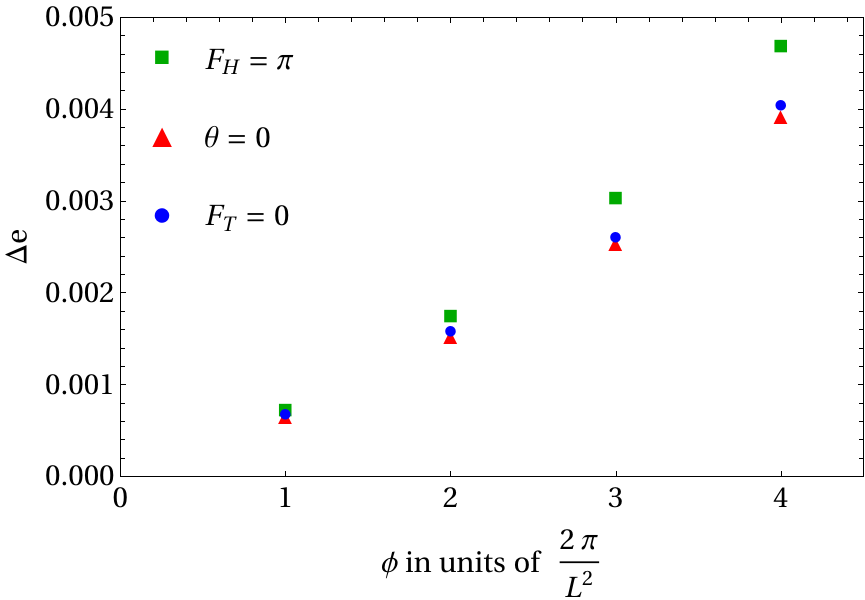}
\caption{\label{fig:twocuts} 
Energy (per site) difference between chiral and Dirac states as a function of $\phi$ for three cuts in the plane of Fig.~\ref{fig:fluxes}. 
Variational Monte Carlo calculations are performed on a cluster with $L=8$. The values of $\phi$ correspond to $N_{\rm mp}=1,\dots,4$ monopoles 
in the torus.}
\end{figure}

\begin{figure}
\includegraphics[width=\columnwidth]{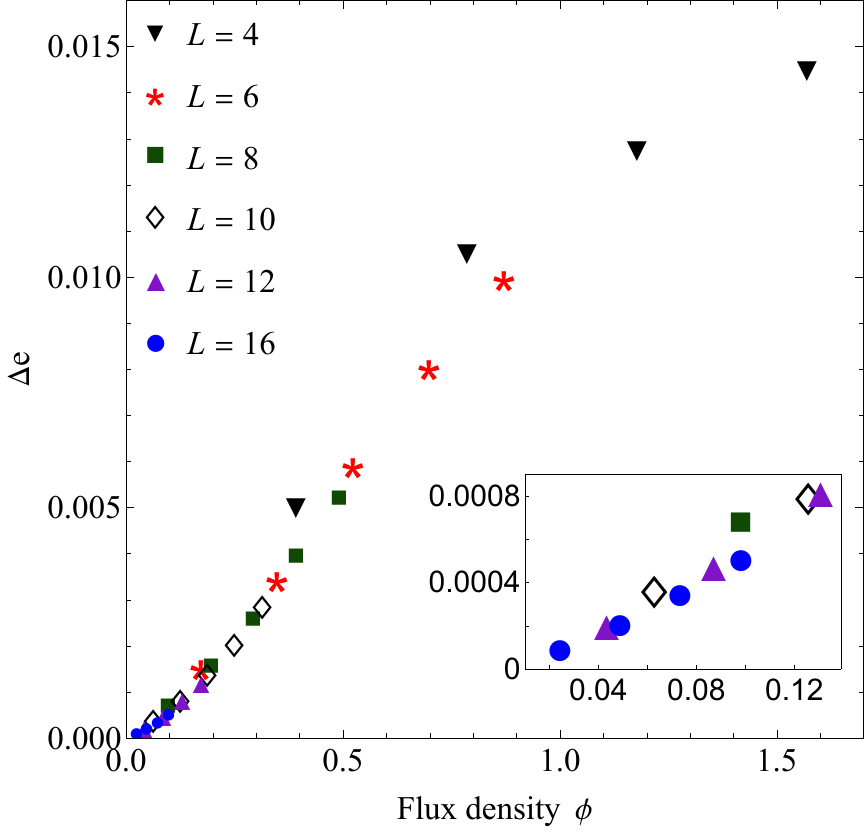}
\caption{\label{fig:phi0} 
Energy (per site) difference between chiral and Dirac states as a function of $\phi$ for $\theta=0$, i.e., the x-axis of the plane shown in 
Fig.~\ref{fig:fluxes}. The variational Monte Carlo calculations are done for both commensurate and monopole fluxes. Inset: zoom of the results 
for small values of $\phi$, where only monopole configurations are present.}
\end{figure}

{\it Results.}
The main outcome of this Letter is that the Dirac state is stable when considering fluxes $\phi \ne 0$ and/or $\theta \ne 0$. Indeed, the best
variational energy (per site) when varying $\theta$ and $\phi$ is obtained for $\theta=\phi=0$, corresponding to the $[\pi,0]$ case. As an 
example, in Fig.~\ref{fig:twocuts}, the variational energies for different cuts in the $(\phi,\theta)$ plane are reported for $L=8$: along 
$\theta=3\phi/8$ (i.e., $F_{H}=\pi$, which connects the Dirac state to the $[\pi,\pi]$ one), along $\theta=-\phi/8$ (i.e., $F_{T}=0$, which 
connects the Dirac state to the $[0,0]$ one), and $\theta=0$. In all cases, the energy increases with $\phi$, even for the smallest possible 
values obtained with a few monopoles. Similar results have been obtained for larger cluster sizes and different cuts. In particular, the case 
with $\theta=0$ is reported in Fig.~\ref{fig:phi0}, where several sizes of the cluster are reported from $L=4$ to $L=16$, including both 
commensurate fluxes (the smallest one being $\phi=2\pi/L$) and monopole configurations (which allow us to reach much smaller values of the 
fluxes). Our results clearly show that the minimal variational energy is always obtained with $\phi=0$, i.e., for the Dirac state.

\begin{figure}
\includegraphics[width=\columnwidth]{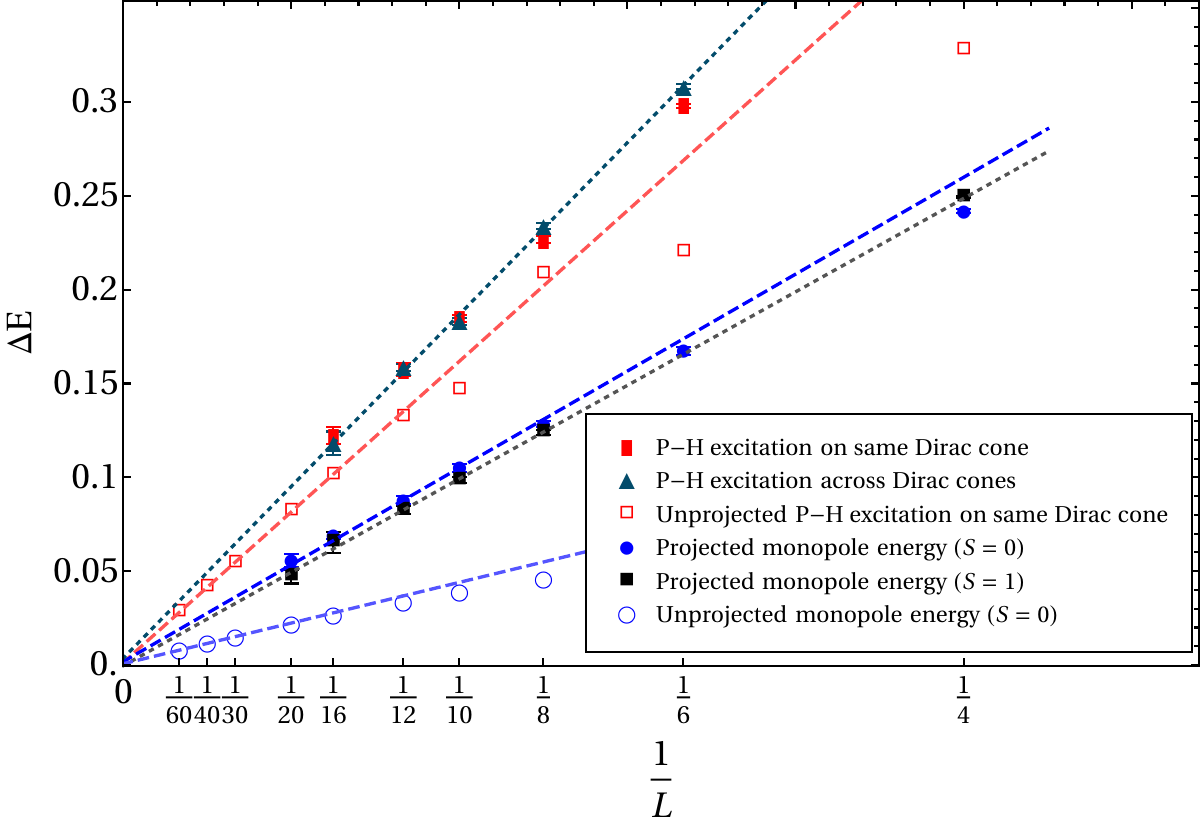}
\caption{\label{fig:scaling} 
Size scaling of the single-monopole gap (with respect to the Dirac state), both singlet and triplet cases are shown. The unprojected case (no
Gutzwiller projection) is reported for comparison. Particle-hole (P-H) spinon excitations of the Dirac wave function are also shown, either within 
the same Dirac cone or across the Dirac cones. }
\end{figure}

Next, we perform the explicit size-scaling analysis of the single-monopole gap, see Fig.~\ref{fig:scaling}. At the unprojected level, i.e., 
when the Gutzwiller projection of Eq.~\eqref{eq:gutz} is not imposed, the monopole configugration corresponds to an excited state that becomes
gapless in the thermodynamic limit. Obviously, this result does not depend on the filling of the degenerate levels at the Fermi level, including 
the case where a triplet state is taken. We emphasize that the vanishing extrapolation becomes evident only when large clusters are considered 
(e.g., $L \gtrsim 30$), since a fitting procedure that only includes $L \lesssim 12$ would predict a finite gap for $L \to \infty$. Most 
importantly, the presence of the Gutzwiller projection has no effect on the overall behavior. In fact, while the slope of the fit is increased, 
the extrapolated value in the thermodynamic limit is always consistent (within a few errorbars) with a vanishing gap. In addition, there is no
appreciable difference (for large clusters) between states with $S=0$ (two fermions occupying orbitals at the Fermi level with up and down 
spins) or $S=1$ (two fermions occupying the orbitals with the same spin). Note that, more generally, monopole excitations in the SU(N$_f$) 
Heisenberg model~\cite{affleck1988b} with N$_f$ even and N$_f/2>1$ fermions per site were also found to be gapless~\cite{SUN}. 

In order to prove (and improve) the statement that spinons are gapless, we construct particle-hole excitations of the Hamiltonian~(\ref{eq:aux_ham}),
by changing the fermion occupation in the unprojected state (i.e., by emptying one of the highest-energy single-particle orbital and filling one 
of the lowest-energy ones). Given the shape of the cluster, there are several ways to do this, since both these shells are four-fold degenerate 
(for each spin value). In particular, we can perform excitations within the same Dirac cone or across the two cones. Trivially, these states are 
gapless in the unprojected wave function, when $L \to \infty$. Most interestingly, they remain gapless even when the Gutzwiller projection is 
included. As a consequence, the $[\pi,0]$ {\it Ansatz}, obtained from the auxiliary Hamiltonian~(\ref{eq:aux_ham}) with {\it real} hoppings 
$\chi^\dagga_{i,j} = \pm 1$, has the remarkable property to describe the (approximated) ground-state wave function that sustain gapless excitation 
for both spinons~\cite{iqbal2014} and monopoles. 

{\it Discussion.} 
In this Letter, we constructed monopole excitations on top of the Dirac spin liquid {\it Ansatz} and showed them to be gapless in the thermodynamic 
limit. By studying the energetics of states with a finite monopole density, we found no sign of an instability towards a chiral state. Our results 
provide further evidence that the ground state of the kagome Heisenberg antiferromagnet is well described by the Dirac spin liquid, despite having 
gapless monopole excitations~\cite{song2019}. Such a remarkable robustness was recently linked to free-fermion band topology dictating symmetry 
properties of monopoles~\cite{song2020}. Recently, a similar analysis of monopole and bilinear excitations was performed on the Dirac spin 
liquid on the triangular lattice~\cite{wietek2023}.
The existence of gapless monopoles may provide new experimental ways to identify U(1) Dirac spin liquids and, in particular, to resolve between gapless $Z_2$ and U(1) states. Recently, a few possibilities have been suggested, e.g., via recently proposed "monopole Josephson effect" \cite{monopolejosephson}, which would lead to a measurable spin current, or via the coupling between monopoles and phonons \cite{seifert2023spinpeierls}, which would lead to a broadening/softening of certain phonon modes.

{\it Acknowledgements.} We thank L. Di Pietro, A. L\"auchli, U. Seifert, C. Wang, J. Knolle, J. Willsher, S. Bhattacharjee, S. Sachdev, S. Capponi, and Y.-C. 
He for helpful discussions. S.~B. also thanks J. Colbois, R. Mishra, and S. Niu for discussions about the project. Y.I., D.P. and S.B. acknowledge 
financial support by the Indo-French Centre for the Promotion of Advanced Research – CEFIPRA Project No. 64T3-1. Y.I. and S.B. would like to 
acknowledge support from the ICTP through the Associates Programme and from the Simons Foundation through grant number 284558FY19, IIT Madras 
through the QuCenDiEM CoE (Project No. SP22231244CPETWOQCDHOC), the International Centre for Theoretical Sciences (ICTS), Bengaluru, India during 
a visit for participating in the program “Frustrated Metals and Insulators” (Code: ICTS/frumi2022/9). The research of Y.I. was supported 
in part by the National Science Foundation under Grant No.~NSF~PHY-1748958. The work of Y.I. was performed in part and completed at the Aspen 
Center for Physics, which is supported by National Science Foundation grant PHY-2210452. The participation of Y.I. at the Aspen Center for Physics 
was supported by the Simons Foundation. Y.I. and S.B. acknowledge the use of the computing resources at HPCE, IIT Madras. This work was granted 
access to the HPC resources of CALMIP center under the allocation 2017-P1231. This work was also supported by the TNTOP ANR-18-CE30-0026-01 grant 
awarded by the French Research Council.

%


\newcommand{\beginsupplement}{%
        \setcounter{table}{0}
        \renewcommand{\thetable}{S\arabic{table}}
        \renewcommand{\theHtable}{S\arabic{table}}%
        \setcounter{figure}{0}
        \renewcommand{\thefigure}{S\arabic{figure}}%
        \renewcommand{\theHfigure}{S\arabic{figure}}
        \setcounter{equation}{0}
        \renewcommand{\theequation}{S\arabic{equation}}%
        \renewcommand{\theHequation}{S\arabic{equation}}%
        \setcounter{page}{1}
        \setcounter{secnumdepth}{3}
     }
 
\renewcommand*{\citenumfont}[1]{S#1}
\renewcommand*{\bibnumfmt}[1]{[S#1]}
 
\newcommand\blankpage{%
    \null
    \thispagestyle{empty}%
    \addtocounter{page}{-1}%
    \newpage
}
\newpage
\blankpage
\newpage
\chead{{\large \bf{--- Supplemental Material ---}}}
\thispagestyle{fancy}

\beginsupplement

\maketitle


\section{SU(N$_f$) monopole}
Here, we generalize the investigation described in the main text by studying the behavior of monopole excitations for fermion flavors N$_f>2$. 
For that purpose we consider the following SU(N$_f$) generalization of the Heisenberg Hamiltonian~\cite{affleck1988}
\begin{equation} \label{perm}
{\cal H} = \sum_{\langle i,j \rangle} \sum_{\alpha \beta}^{\rm N_{\it f}} c^\dagger_{i,\alpha} c^\dagga_{i,\beta} c^\dagger_{j,\beta} c^\dagga_{j,\alpha}
\end{equation}
with N$_f$ even integer and N$_f/2$ fermions per site. Here, $\alpha,\beta$ are ``spin'' indices that take the values $\alpha,\beta=1,2,\hdots$N$_f$.
For the standard SU(2) case, this is related to the Heisenberg Hamiltonian as 
\begin{equation} \label{perm_hei}
{\bf S}_i \cdot {\bf S}_j = \frac{1}{2} \sum_{\alpha,\beta} c^\dagger_{i,\alpha} c^\dagga_{i,\beta} c^\dagger_{j,\beta} c_{j,\alpha} - 
\frac{1}{4} n_i n_j
\end{equation}
where 
\begin{equation}
n_i = c^\dagger_{i,\uparrow} c^\dagga_{i,\uparrow} + c^\dagger_{i,\downarrow} c^\dagga_{i\downarrow}\, .
\end{equation}
The latter term in Eq.~\eqref{perm_hei} is just a constant in the subspace with one fermion per site.

\begin{figure}[H]
\includegraphics[width=0.95\columnwidth]{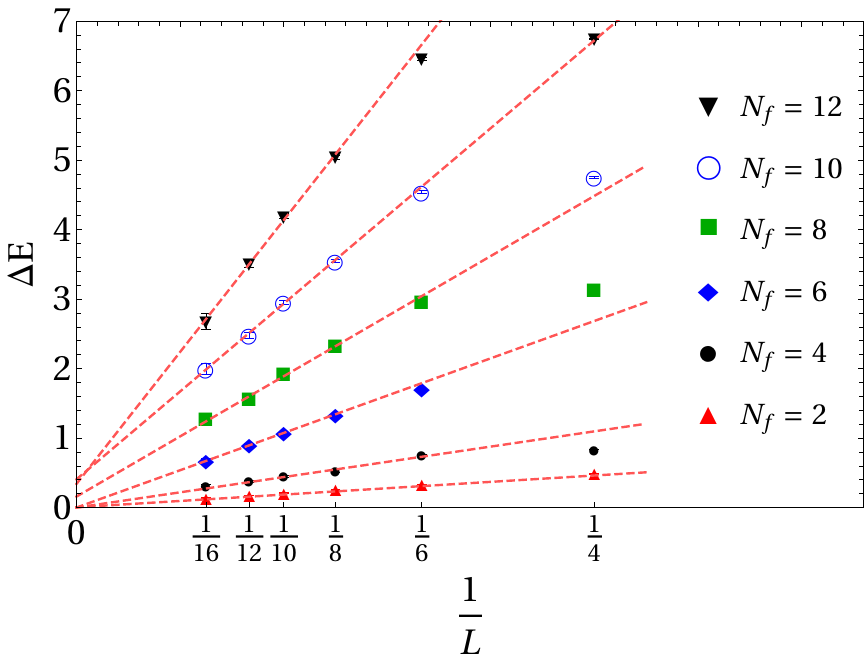}
\caption{\label{fig:sun_monopoles} 
Single-monopole gap as a function of $1/L$ for various values of N$_f$, along with best linear fits.}
\end{figure}

At the unprojected level, the expectation values of~\eqref{perm} can be computed using Wick's theorem. The many-body wave function $\ket{\Phi_0}$ 
is constructed as a product of $N/2$ orbitals for each spin flavor:
\begin{equation}
\ket{\Phi_0} = \prod_{\alpha = 1}^{N_f}   \left(\prod_{x=1}^{N/2} \phi^{\dagger}_{x,\alpha} \right) \ket{0}\, ,
\end{equation}
where $N=3L^2$ is the number of sites. The (orthonormal) orbitals $\phi_{x,\alpha}^\dagger$ are obtained by diagonalizing the relevant free-fermion 
tight-binding model (either the Dirac or the monopole ansatz)
\begin{equation} \label{ferm_orb}
\phi_{x,\alpha}^\dagger = \sum_{j=1}^{N} U_{j,x} c^\dagger_{j,\alpha}\, ,
\end{equation}
where $U$ is the $N \times N$ eigenvector matrix. The unprojected expectation value 
\begin{equation} \label{MF}
E_{0} = \frac{\bra{\Phi_0} H \ket{\Phi_0}}{\bra{\Phi_0}\ket{\Phi_0}} 
\end{equation}
can be evaluated, using Eq.~\eqref{ferm_orb}, as
\begin{equation} \label{fin_result}
E_{0} = - \sum_{\langle i,j \rangle} \left[ {\rm N}_f^2 |A_{i,j}|^2 + \frac{{\rm N}_f}{4} \right]   
\end{equation}
where 
\begin{equation}
A_{i,j} = \sum_{x=1}^{N/2} U_{j,x} U_{i,x}^*\, .
\end{equation}
Thus the coefficients $A_{i,j}$ can be readily calculated from a real-space diagonalization of the tight-binding model. The single-monople gap is
obtained by taking the difference between the case with one monopole (spread over the entire torus) and no monopoles (i.e., the Dirac state):
\begin{equation} \label{final_diff}
\Delta E_{0} = - {\rm N}_f^2  \sum_{\langle i,j \rangle} \left[ |A^{\text{monopole}}_{i,j}|^2 - |A^{\text{Dirac}}_{i,j}|^2  \right]  \, .
\end{equation}

For the projected wave functions, we use the Monte Carlo sampling to evaluate the variational energies corresponding to the single monopole and 
the Dirac state (in both cases, the Gutzwiller projector imposes to have N$_f/2$ fermions per site).

The size scaling of the total monopole energy for N$_f=2,4,6,8,10,$ and $12$ is shown in Fig.~\ref{fig:sun_monopoles}.

\begin{figure}
\includegraphics[width=0.95\columnwidth]{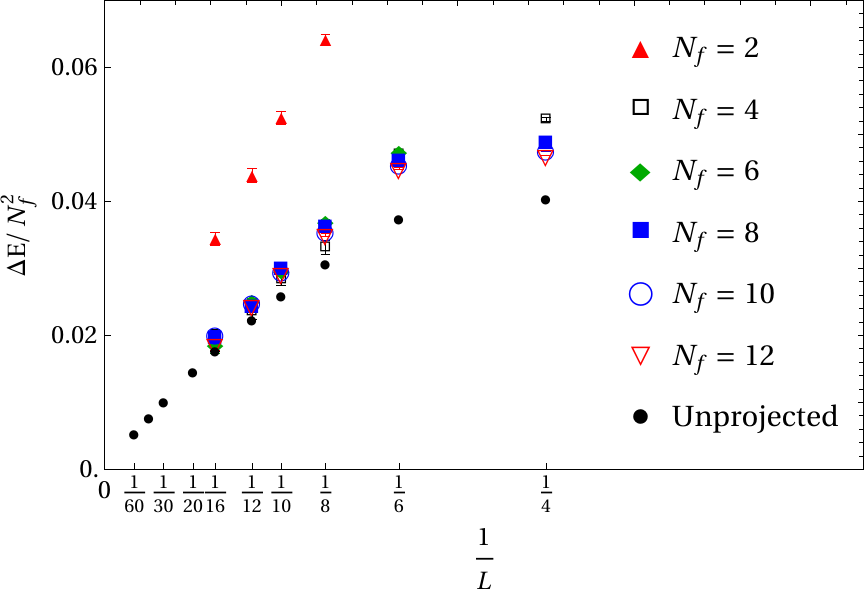}
\caption{\label{fig:sun_monopoles_scaled} 
Single-monopole energies scaled by N$_f^2$, shown along with the unprojected energy difference $\Delta E_{0}/{\rm N}_f^2$ from Eq.~\eqref{final_diff}.} 
\end{figure}

At first glance, the data for N$_f \geq 8$ would suggest a gapped monopole in the thermodynamic limit. However, after further investigation, we 
believe this to be a finite-size effect. Indeed, by scaling the projected energies by N$_f^2$, we observe that all the data (except N$_f=2$) 
collapse perfectly on top of each other, see Fig.~\ref{fig:sun_monopoles_scaled}. Furthermore, there is very good agreement between the projected
and the unprojected energies calculated from Eq.~\eqref{final_diff}, which increases with system size. Finally, we remark that the unprojected 
data indicates a gapless monopole only if large enough system sizes $L \geq 30$ are considered, which are very difficult to access for the 
projected wave functions.

\section{Mean field monopole spectrum and boundary conditions}

In the main text, we remarked that the two-fold degeneracy at the Fermi level upon adding the monopole flux on top of the Dirac state cannot be removed by changing the boundary conditions. Indeed, this is true for both the square and the Kagome lattice. Thus, the open shell at the Fermi level is an unavoidable consequence of the monopole flux. 
We also observed that in general (for large enough lattice sizes $L$), when more monopoles $N_m$ are added, the degeneracy at the fermi level is $2\times N_m$ for small values of $N_m$.
\par
\begin{figure}[H]
\includegraphics[width=0.7\columnwidth]{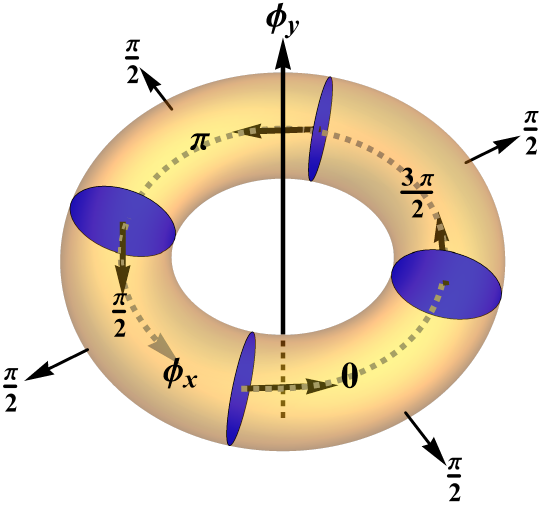}
\caption{\label{fig:torus} 
Fluxes through (one set of) incontractible loops of the $L=4$ torus for the monopole state. Also shown schematically are the angles $\phi_x$ and $\phi_y$, which control boundary conditions.} 
\end{figure}

Since each plaquette has a flux $2\pi/L^2$ piercing it, the total flux through a cylindrical strip of the lattice containing $L$ sites is $2\pi/L$, in contrast to the Landau gauge where it would have been a multiple of $2\pi$. This necessarily implies non-trivial fluxes through the incontractible loops (marked in blue in figure \ref{fig:torus}) of the torus. As a result, translational symmetry is broken along \textit{both} the lattice directions $\Vec{a}_1$ and $\Vec{a}_2$.  

In general, for a $L \times L$ lattice, the fluxes through these loops are $0, 2\pi/L, 4\pi/L, \hdots$.
A natural consequence of these non-trivial fluxes is that the boundary conditions $\phi_x, \phi_y$ can be chosen modulo $2\pi/L$.


\end{document}